\documentclass[12pt]{article}
\def\demi{{\textstyle {1\over2}}}

\def\L{l}

\let\pa=\partial

\let\nonu=\nonumber

\let\nn=\nonumber

\def\V{{\cal V}}
\def\ot{\tilde\o}
\def\tD{\tilde D}

\begin{document}
\bibliographystyle{perso}

\begin{titlepage}
\null \vskip -0.6cm
\hfill PAR--LPTHE 03--11

\hfill hep-th/0303165

\vskip 1.4truecm
\begin{center}
\obeylines
        {\Large
        Supergravity and the Knitting of the Kalb--Ramond
        Two-Form in Eight-Dimensional Topological Gravity.
\vskip 6mm}
Laurent Baulieu, Marc Bellon,  Alessandro Tanzini
{\em Laboratoire de Physique Th\'eorique et Hautes Energies,
 Universit\'es Pierre et Marie Curie, Paris~VI
et Denis~Diderot,~Paris~VII}

\end{center}

\vskip 13mm

\noindent{\bf Abstract}:  Topological euclidean gravity is built 
in eight dimensions for manifolds with
$Spin(7) \subset SO(8)$ holonomy. In a previous work, we
considered the construction of an eight-dimensional 
topological theory describing the graviton and one  graviphoton. 
Here we solve the question of determining a
topological model for the combined system of a metric and a 
Kalb--Ramond two-form gauge field. We then  recover the complete 
$N=1, D=8$ supergravity theory in a twisted form. 
We observe that the generalized
self-duality conditions of our model correspond to
the octonionic string equations.
\vfill

\begin{center}

\hrule \medskip
\obeylines
Postal address: %
Laboratoire de Physique Th\'eorique et des Hautes Energies,
 Unit\'e Mixte de Recherche CNRS 7589,
 Universit\'e Pierre et Marie Curie, bo\^\i te postale 126.
4, place Jussieu, F--75252 PARIS Cedex 05

\end{center}
\end{titlepage}
\def\w{\wedge}
\def\o{\omega}
\def\t{\tilde}
\def\TQFT{ Topological Field Theory}
\section{Introduction}

All types of
superstring theories can be formally obtained by suitable anomaly free
untwisting of a topological sigma-model \cite{green}. This suggests the
possibility that  supergravities, which arise as
low energy limits of superstrings, can be understood as 
topological gravities. In particular,
$D=11$ supergravity, which   determines all known supergravities in lower
dimensions,  could   be viewed as a  topological theory. 

In a previous work we have shown that, both in four \cite{BT1} and in
eight dimensions \cite{BT2}, the   Einstein action plus the
Rarita--Schwinger term (in a twisted form) can  be obtained  by
constructing a topological quantum field theory (TQFT), which
implements in a BRST invariant way the gravitational instanton
equation.\footnote{In four dimensions, the twist of the complete action
including the interaction terms has been discussed in \cite{mespe}.}
These constructions only holds for  manifolds with special holonomy,
{\em i.e.}, $SU(2)\subset SO(4)$ in four dimensions and $Spin(7)\subset
SO(8)$ in eight dimensions.

In \cite{BT2}, we left open the delicate point of 
introducing  a sector of the eight-dimensional TQFT involving a two-form
gauge field. This case escapes the
procedure displayed in \cite {laroche}. 
In this paper, the equations for the two-form gauge field
appear mixed with the one for other fields. 
The basic idea is the construction of a TQFT for a general two-tensor
field, which is the natural object stemming from the
zero slope limit of topological sigma-models. The symmetric part
of this tensor describes the metric,
while its antisymmetric part gives the two-form field.
We use the formalism where the metric is described as 
a vielbein modulo the Lorentz symmetry.
This means adding and subtracting spurious
degrees of freedom, a task which is by now familiar in the context of
TQFT's. 
As in our previous model \cite{BT2}, 
we find that the topological theory is defined on manifolds with
$Spin(7)$ holonomy. In fact, the $Spin(7)$-invariant four-form
plays a central r\^ole in the determination of the topological gauge functions. 
Moreover, the presence on these manifolds of a covariantly constant
spinor allows for the definition of a twist \cite{BT2}
which maps some of the fermionic ghosts and antighosts of the topological
model on the spinors of $N=1, D=8$ supergravity.
Finally, some ghosts of ghost and ghosts of antighosts can be untwisted
into the commuting ghosts of local supersymmetry, explaining the emergence
of local supersymmetry in our model.

\let\w=\wedge
\let\o=\omega
\let\O=\Omega
\def\L{ {\cal L}}
 \section{Including the two-form in topological gravity }  

In this section we address the question of building a TQFT multiplet
for a general tensor $A_{\mu\nu}$ of rank two.  We consider an
eight-dimensional manifold with holonomy group $Spin(7)\subset SO(8)$.
The tensor $A_{\mu\nu}$ can be split into its symmetric and
antisymmetric parts.  If, as in \cite{BT2},  we  only consider  the
symmetric part, which can be interpreted as a metric, we can construct
a TQFT that contains the Einstein action, by using a gravitational
octonionic self-duality equation. In the spirit of \cite{BT1} we can
also introduce a coupling to the TQFT for an abelian graviphoton, which
has ghost number two. This topological model determines, in a twisted
form, a truncation of $N=1,D=8$ supergravity \cite{BT2}.  However, the
dilaton, the Kalb--Ramond two-form, one graviphoton and their fermionic
superpartners in the $N=1,D=8$ supergravity multiplet \cite{sase}
escape this construction.

The  difficulty of determining a TQFT for the antisymmetric part
$B_{\mu\nu}$ of $A_{\mu\nu}$ can be appreciated as follows, following the ideas
contained in \cite{laroche}. The two-form $B_{\mu\nu}$ contains 28 components,
which  give 21 degrees of freedom modulo the gauge invariance  
$B_{\mu\nu}\sim B_{\mu\nu}+\partial_{[\mu}\Lambda_{\nu]}$.
The field strength of $B_{\mu\nu}$ is a three-form $G_3=dB_2$ containing
$56=(^8_3)$ components. The analysis in~\cite{laroche} indicates
that there is no natural way to chose an holonomy group for
the eight-dimensional manifold which allows to 
write a self-duality equation for $G_3$, which, (i)  
would count for 21 topological independent equations and, (ii),
would solve the relativistic wave equation of a two-form.
Moreover, the 28 components of the topological ghost of
a two-form cannot be rearranged in eight-dimensional spinor
representations. 
Rather, to determine a TQFT involving the two-form 
in eight dimensions, we will see that it is
necessary to combine the Lorentz invariance and the topological 
invariance of the two-form, and write self-duality equations that mix the
dilaton, the eight-bein and the two-form.  That this is not only possible,
but a useful complement of the construction of~\cite{BT2} appears
immediately from the fact that there were 56 components for the antighost
for the vielbein, more than the 28 degrees of freedom it describes.

To start with, we enlarge the question of building a TQFT for 
$A_{\mu\nu}$ into that of building a TQFT for a vielbein $e^a_\mu$,  a Lorentz
ghost $\Omega^{ab} $ and  a two-form   $B_{\mu\nu}$,%
\footnote{Throughout the paper the Latin indices
$a,b,\ldots$ denote flat $SO(8)$ tangent space indices, and
$\mu,\nu,\ldots$ are eight-dimensional world indices.}
$$A_{\mu\nu}= A_{\{\mu\nu\}} \oplus  A_{[\mu\nu]} \to (e^a_\mu,
\Omega^{ab},B_{\mu\nu}). $$
The Lorentz symmetry can be used to set to zero the
antisymmetric part of the matrix $e^a_\mu$\footnote{Strictly speaking,
the two indices of the vielbein represent components in different spaces,
so that one can only speak of the antisymmetric part of $e_\mu^a$ once a
background vielbein has been chosen. This is not the case for other fields
with the same indices as the vielbein, since the vielbein is then available
to relate the two spaces.}.
The number of degrees of freedom of the system 
$A_{\mu\nu}=A_{[\mu\nu]} \oplus A_{\{\mu\nu\}}$  is indeed
equal to that of the $(e^a_\mu, B_{\mu\nu} ,\Omega^{ab})$
system, when the components are
algebraically counted, since the Lorentz ghost $\Omega^{ab}$ counts
negatively. Eventually, we will interpret 
$A_{\{\mu\nu\}}$ as a metric $g_{\mu\nu}$,
assuming that $e^a_\mu$ is an invertible matrix.

In contrast with our previous work, graviphotons   are not separately
introduced. Here, these fields   naturally appear  as ghosts for the
topological ghosts of the two-form $B_{\mu\nu}$. An analogous situation
holds for the dilaton, which is a $Spin(7)$-invariant part of a ghost
of ghost for the vielbein. Having obtained a truncation of supergravity
in \cite{BT2} is now understood  as having  consistently retained a
part of the topological BRST multiplet.

Eventually, we will recognize that we have a TQFT with an equivariance
with respect  to the Lorentz$\times$diffeomorphism symmetry, whose
gauge fields are the spin connection $\o^{ab}_\mu$ and the vielbein
$e^a_\mu$.  Moreover, our topological model displays an equivariance
with respect to the vector gauge symmetry of the two-form, $B_2\sim B_2
+ d\Lambda_1$.  Finally, local supersymmetry will show up  as a
consequence of the symmetry of the topological ghost of the  vielbein,
defined modulo reparametrizations.  Shortly speaking, the construction
of a TQFT for a two-tensor yields all the  fields of supergravity. As
we will discuss in the next section, the topological gauge functions
are given by self-duality equations which mix the symmetric and
antisymmetric parts of the two-tensor.

Let us now proceed to the detailed construction of the BRST
topological multiplets. Geometry determines the set of ghosts and our first
guess for the complete set of fields for an $SO(8)$ invariant  TQFT  
with a vielbein $e^a_\mu$, a two-form $B_{\mu\nu}$ and  a spin connection
$ \o^{ab}_\mu$ is:
\let\hw=\hidewidth
\begin{eqnarray}
\matrix
{  &    &  e^a_\mu  &   & \cr
   &\Psi^{(1) a}_\mu  &    &  \bar\Psi^{(-1)a }_\mu     & \cr
   \Phi^{(2)a} &  &   \Phi^{(0)a} ,  b^{(0)a }_\mu &  & \bar\Phi^{(-2)a}\cr
   &  \eta^{(1 )a} &     & \bar \eta^{(-1 )a} & \cr }
\nonu
\end{eqnarray}

\begin{eqnarray}
\matrix
{  &    &  \o^{ab}_\mu  &   & \cr
&  \tilde\Psi^{(1)ab}_\mu  &    & \bar{\tilde\Psi}^{(-1)ab}_\mu & \cr
{\tilde \Phi}^{(2)ab } &    &  {\tilde \Phi}^{(0)ab},{\tilde b}^{(0)ab}_\mu
   &   & \bar{\tilde\Phi}^{(-2)ab}  \cr
&\hw {\tilde\eta}^{(1)ab } \hw&   &  \hw\bar {\tilde\eta}^{{(-1)ab} } \hw& \cr }
\nonu
\end{eqnarray}

\begin{eqnarray}
\matrix
{  &    &    &      B_{\mu \nu} &   &   &   \cr
   &    & \Psi^{(1)}_{\mu \nu}  &   &   \bar \Psi^{(-1)}_{\mu\nu}  &   \cr
   & A^{(2)}_{\mu } &    &  A^{(0)}_{\mu } , b^{(0)}_{\mu \nu}  &&   
  A^{(-2)}_{\mu }  &  \cr
  R^{(3)} &  &  S^{(1)},\Psi^{(1)}_\mu && \bar S^{(-1)},\bar \Psi^{(-1)}_\mu &
  &   \bar R^{(-3)}  \cr
  & \hw b_ {S^{(1)}} ^{2},\Phi^{(2)} \hw&&\hw  b_ {\bar S^{(-1)}} ^{(0)},\bar
\Phi^{(0)},\Phi^{(0)} \hw&  &\hw b_{\bar R^{(-3)}}^{(-2)},\bar \Phi^{(-2)}
\hw&  \cr
 &&          \eta^{(1)} && \bar \eta^{(-1)}      &   \cr}
\nonu
\end{eqnarray}
\begin{eqnarray}
\matrix
{ \xi^{(1) \mu}   &     &    \bar \xi^{(-1) \mu} \cr
     &    b^{(0) \mu} }
\quad\quad\quad\quad\quad
\matrix
{ \O^ {(1)ab}  &   &        \bar \O^ {(-1)ab} \cr
     &  b^{(0)ab}  & }
\end{eqnarray}
For the sake of clarity, we have made explicit (as an upper index) 
the ghost number of the fields in the ``pyramid"
that describes the BRST topological multiplets. 
We could introduce a bigrading that separate the ghost
number and antighost number, but this would make heavier the
notations. In the above pyramids, the BRST symmetry
acts on the south-west direction. The fields which are not 
on the left edge of each pyramid are topological pairs made of  
antighosts and  their Lagrange
multipliers. They   satisfy trivial BRST equations\footnote{More
precisely, all equations for the antighosts appearing in
our field spectrum are of the type
\def\s{\hat s}
\def\bg{\bar  g}
$ \s \bg = \lambda $, $  \s \lambda =\L_\Phi \bg + \delta_{\t \Phi}\bg $
with  $s X= \s X +\L_\xi X+\delta_{{\O} } X$.}. Actually, each one of
the  fields that are labeled by a letter  $b$ or 
$\eta$, with various indices, is a bosonic or fermionic Lagrange
multiplier  field, and is essentially equal to the BRST variation of
the antighost that is located at its  upper right  position, e.g., $s
\Psi^{(-1)}_{\mu\nu}=b^{(0)}_{\mu\nu}+\ldots$,
$s\bar S^{(-1)}=  b_ {\bar S^{(-1)}} ^{(0)}+\ldots$. As an
exception to this notational rule, we find useful to define the fields
$\Psi^{(1)}_\mu$ and $\bar \Psi^{(-1)}_\mu$ as the   fermionic Lagrange
multipliers that  stem  from the BRST variation  of  the
commuting ghosts of ghosts
$A^{(0)}_\mu$ and $A^{(-2)}_\mu$ respectively, 
{\it i.e.} $sA^{(0)}_\mu= \Psi^{(1)}_\mu+\ldots$
and $sA^{(-2)}_\mu= \bar\Psi^{(-1)}_\mu+\ldots$.

The fields which carry  the essential geometrical information are on the
left edge of the pyramids. Their topological symmetry is defined
as:
\begin{eqnarray}
se^a_\mu &=& \Psi^{(1)a}_\mu-\O^{ab} e^b_\mu  + \L_\xi e^a_\mu ,\nonumber
\\ s\o^{ab}_\mu &=&   \t \Psi^{(1)ab}_\mu+
D_\mu\O^{ab}  + \L_\xi \o^{ab}_\mu, 
\nonumber\\
\nonumber\\ 
s\Psi^{(1)a}_\mu &=&  -\O^{ab} \Psi^{(1)b}_\mu-\L_\Phi
e^a_\mu +\t \Phi^{(2)ab}e^b_\mu +\L_\xi  \Psi^a_\mu,
\nonumber\\
s \t \Psi ^{(2)ab}_\mu &=& -\O ^{ac}\t
\Psi^{(2)cb}  + D_\mu \t \Phi^{(2)ab} -\L _\Phi \o^{ab}_\mu 
+ \L_\xi  \t\Psi ^{(2)ab}_\mu,
\nonumber\\
\nonumber\\
s  \Phi ^{(2)a}  &=& \L_\xi   \Phi ^{(2)a } -\O ^{ac} \Phi^{(2)a}  ,
\nonumber\\
s \t \Phi ^{(2)ab}  &=&  -\O ^{ac}\t \Phi^{(2)cb}  
+\L_\xi  \t\Phi ^{(2)ab},
\nonumber\\
\nonumber\\
sB_{\mu\nu} &=& 
\Psi^{(1)}_{\mu\nu}
+\L_\xi B_{\mu\nu},
\nonumber \\ 
s\Psi^{(1)}_{\mu\nu} &=&\L_\Phi B_{\mu\nu}
 +\partial_{[\mu} A^{(2)}_{ \nu]} + \L_\xi \Psi^{(1)}_{\mu\nu},
\nonumber \\ 
sA^{(2)}_{\mu } &=& \partial_{\mu} R^{(3)} + \L_\xi A^{(2)}_{\mu } ,
\nonumber \\ 
s R^{(3)} &=& \L_\xi   R^{(3)} ,
\nonumber\\
\nonumber\\ 
s\xi^\mu  &=& f_a^\mu \Phi^{(2)a}+\xi^\nu \partial _\nu \xi^\mu,
\nonumber\\
s \O ^{ab} &=& \t \Phi^{(2)ab} -\O ^{ac}\O^{cb}  + \L_\xi  \O ^{ab}  .
\label{brs} 
\end{eqnarray}
In the variation of $\xi^\mu$, the inverse of the vielbein $e_\mu^a$
appears and we denote it as $f_a^\mu $.
We have  not introduced   ghosts $V_\mu^{(1) }$ and ghost of ghost $ 
m^{(2)}$ for    the gauge invariance of the two-form gauge field, with the
standard BRST symmetry
$QB_{\mu\nu}=\partial _{[\mu} V^{(1) }_{\nu]}$,
$QV^{(1) }_{\mu}=\partial _{ \mu} m^{(2) }$, $Qm^{(2) }=0$, which would
yield   $s V^{(1)}_\mu=A^{(2)}_\mu$ and
$sm^{(2)}= R^{(3)}$.  
We choose instead to write equivariant BRST transformations 
with respect to this symmetry. This implies that 
the square of the BRST transformations on the field 
$B_{\mu\nu}$ is not zero, but corresponds to a
reparametrization with parameter $\Phi^a$
and an abelian transformation $B_{\mu\nu}\to B_{\mu\nu}+\partial
_{[\mu} \Lambda_{\nu]}$, with $\Lambda_\mu=A^{(2) }_\mu$. 
The BRST operator $s$ is thus nilpotent only 
modulo gauge transformations for the two-form gauge field.
Actually, the topological
gauge functions we will use in sect.3
only involve the curvature $G_3=dB_2$, so there
is no need to give the details of the gauge symmetry of the two-form.
We could further set $\xi^\mu=0$, which would yield a BRST symmetry equivariant
with respect to the reparametrization. In this case, 
the BRST symmetry would be nilpotent also modulo
reparametrizations along the vector ghost of ghosts $ e_a^\mu \Phi^{(2)a}$. 
As for the Lorentz invariance, we will instead carefully keep the Lorentz ghost
dependence. Eventually, the corresponding gauge functions 
will be equivariant with respect to local Lorentz transformations 
for $Spin(7)\subset SO(8)$.

There is an $U(1)$ invariance 
for the fields $\Psi^{(1)}_\mu$ and 
$\bar \Psi^{(-1)}_\mu$. 
Also for this invariance we prefer to work 
with an equivariant BRST operator, and we do not
write explicitly the related Faddeev--Popov ghosts 
$sc =\Phi^{(2)}$, $s\bar c =\bar \Phi^{(0)}$ in the BRST transformations
\begin{eqnarray}
s  A^{(-2)}_\mu &=&  \bar \Psi^{(-1)}_\mu +\L_\xi  A^{(-2)}_\mu
\ \ \ \ \ \ s  A^{(0)}_\mu = \Psi^{(1)}_\mu +\L_\xi  A^{(0)}_\mu
 \nonumber \\
s\bar \Psi^{(-1)}_\mu &=& \partial  _\mu  \Phi^{(0)}
+ \L_\xi \bar \Psi^{-1}_\mu 
\ \ \ \ \ \ \ 
s \Psi^{(1)}_\mu = \partial  _\mu   \Phi^{(2)} +\L_\xi   \Psi^{(1)}_\mu 
 \nonumber \\
s  \Phi^{(0)} &=& \L_\xi  \Phi^{(0)}
\ \ \ \ \ \ \  \ \ \ \ \ \ \ \ \
\ \ \ s  \Phi^{(2)} = \L_\xi   \Phi^{(2)}
 \nonumber \\
s \bar \Phi^{(0)} &=& \L_\xi \bar \Phi^{(0)}  +\eta^{(1)}
\ \ \ \ \  \ \ \ \ \
s\bar \Phi^{(-2)}= \bar \eta^{{(-1)}}+ \L_\xi \bar \Phi^ {{(-2)}}
 \nonumber \\
s   \eta^{(1)} &=& \L_\xi  \eta^{(1)}
\ \ \ \ \ \ \  \ \ \ \ \ \ \ \ \ \ \ \ \
s  \bar \eta^{{(-1)}}= \L_\xi \bar \eta^{{(-1)}} \ .
\end{eqnarray}
If we only retain a $Spin(7)\subset SO(8)$ invariance, we can
redistribute the degrees of freedom of the antighosts and Lagrange
multipliers  of the BRST  multiplets. 
In this way, we shall be able to define
$Spin(7)\subset SO(8)$ invariant topological gauge
functions, which is the key for building the eight-dimensional TQFT.
Using the decomposition of a
$SO(8)$ valued two-form $M^{ab }=M^{ab^-}+ M^{ab^+} $ 
in the $Spin(7)$-invariant representations $28=7\oplus21$, we get
\begin{eqnarray}
\matrix
{  &     &  e^a_\mu    &   & \cr
    &  \Psi^{(1)a}_\mu  &    &  \bar\Psi^{(-1)ab^-}_\mu,\bar\chi^{(-1)a} & \cr
    \Phi^{(2)a}   & & \sigma,  \Phi^{(0)ab^-}\; b^{(0)ab^-}_\mu,  b^{(0) }_\mu 
   & &\bar\Phi^{(-2)a} \cr
   &\chi^{( 1 )},  \eta^{(1 )ab^-  }  & &    \bar \eta^{(-1)a } & \cr
  }
\nonu
\end{eqnarray}

\begin{eqnarray}
\matrix
{  &    &  \o^{ab}_\mu  &   &  \cr
 &  \tilde\Psi^{(1)ab}_\mu    & & \bar{\tilde\Psi}^{(-1)ab}_\mu & \cr
     { \tilde\Phi}^{(2)ab ^\pm }   &    &  { \tilde\Phi}^{(0)ab ^\pm },  
 {\tilde b}^{(0)ab}_\mu &   & {\bar{\tilde\Phi}}^{(-2)ab ^\pm } \cr
 &   { \tilde\eta}^{(1)ab ^\pm }  & & {\bar{\tilde\eta}}^{(-1)ab ^\pm } & \cr
  }
\nonu
\end{eqnarray}

\begin{eqnarray}
\matrix
{  &  &  &     B_{\mu \nu} &   &   &   \cr
   &  &   \Psi^{(1)}_{\mu \nu}  &     &  \bar \Psi^{(-1) }_{\mu\nu^\pm} && \cr
   & A^{(2)}_{\mu } &&  A^{(0)},  A^{(0)}_{\mu \nu^-}, b^{(0)  }_{\mu \nu^\pm}
   & &   A^{(-2)}_{\mu } &  \cr
   R^{(3)} &   &  \hw S^{(1)},\Psi^{(1)}_{\mu\nu^-}, \Psi^{(1)}\hw & & 
\hw \bar S^{(-1)},\bar \Psi^{(-1)}_{\mu} \hw & &    \bar R^{(-3)}  \cr
    &   b_ {S^{(1)}} ^{(2)},   \Phi^{(2)} &  & b_ {\bar S^{(-1)}} ^{(0)},
\Phi^{(0)} ,\bar\Phi ^{(0)}  & &      b_{\bar R^{(-3)}}^{(-2)},\bar \Phi^{(-2)}
&    \cr
       & & \eta^{(1)} &  &  \eta^{(-1)}&     &   \cr
}
\nonu
\end{eqnarray}

\begin{eqnarray}
\matrix
{ \xi^ \mu   &     &    \bar \xi^ \mu \cr
     &    b^\mu }
\quad\quad\quad\quad\quad
\matrix
{
      \O^ {ab ^\pm}  &   &        \bar \O^ {ab ^ \pm} \cr
     &  b^{ab ^\pm}  &
  }
\end{eqnarray}
Let us now clarify the r\^ole of the various
topological fields and their
supergravity interpretation.
To obtain the relevant set of propagating fields, a certain number
of trivial gauge fixing must be done, which give  algebraic terms of the
type
$s(AB)=(sA)B\pm A(sB)=\Psi_A A\pm A\Psi_B$. Such terms  
eliminate quartets of the form
$A,\Psi_A, B,\Psi _B$ in a BRST invariant  way by their algebraic
equations of motions.
After some thinking, one understands that the
following $s$-exact terms are needed
\begin{eqnarray}
&&s\Big[{\bar{\tilde \Phi}} ^{(-2 )ab^-} 
( \t \eta  ^{(1 )ab^-}\,\, -\O  ^{ ab^-}) \Big]=
 {\bar{\tilde \eta}} ^{(-1 )ab^-}
( {\t \eta}  ^{(1 )ab^-}\,\, - \O  ^{ ab^-})
+ {\bar{\tilde \Phi}} ^{(-2 )ab^-} {\tilde \Phi}^{(2 )ab^-} 
\cr
&&s\Big[{\bar{ \Psi}} ^{(-1 ) }_{\mu\nu ^+ }
{ {\tilde \Phi}} ^{(0 )ab^+} e^\mu_ae^\nu_b
\Big]
={\bar{ \Psi}} ^{(-1 ) }_{\mu\nu ^+ }
{ {\tilde \eta}} ^{(1 )ab^+} e^\mu_ae^\nu_b
+{ {b}} ^{(0 ) }_{\mu\nu ^+ }
{ {\tilde \Phi}} ^{(0 )ab^+} e^\mu_ae^\nu_b
\cr
&&s\Big[{\bar{ \Psi}} ^{(-1 ) }_{\mu\nu ^- } { {A}} ^{(0 ) }_{\mu\nu^-} \Big]
= {\bar{ \Psi}} ^{(-1 ) }_{\mu\nu ^- } { {\Psi}} ^{(1 ) }_{\mu\nu^-}
+ {\bar{ b}} ^{(0 ) }_{\mu\nu ^- } { {A}} ^{(0 ) }_{\mu\nu^-}
\nonumber \\
&&s \Big[{\bar S}^{(-1)}A^{(0)} \Big] = \bar S^{(-1)}\Psi^{(1)} 
+b _{\bar S^{(-1)}}^{(0)}  A^{(0)}
\end{eqnarray}
The fields $\Phi^{(2)}, \bar \Phi^{(-2)},\bar \eta^{(-1)}$ are also
eliminated, together with the quartets for  $A^{(0)}_{\mu\nu}$. 
The remaining fields are then:
\begin{eqnarray}
\matrix
{   &    &  e^a_\mu   &   & \cr
    &  \Psi^{(1)a}_\mu  &    &\hw  \bar\Psi^{(-1)ab^-}_\mu,
            \bar\chi^{(-1)a} \hw& \cr
    \Phi^{(2)a}  &    & \sigma,  \Phi^{(0)ab^-}\; b^{(0)ab^-}_\mu,b^{(0) }_\mu
    & & \bar\Phi^{(-2)a} \cr
    &\hw   \chi^{( 1) },  \eta^{(1) ab^-  } \hw&&    \bar \eta^{(-1)a } & \cr
  }
\nonu
\end{eqnarray}

\begin{eqnarray}
\matrix
{   &    &  \o^{ab}_\mu  &   &   \cr
    &  \tilde\Psi^{1ab}_\mu   & & \bar{\tilde\Psi}^{-1ab}_\mu &  \cr
    { \tilde\Phi}^{2ab ^+ }    &&     \t\Phi^{(0)ab^-}, {\tilde b}^{0ab}_\mu 
           &   & {\bar{\tilde\Phi}}^{-2ab ^+} \cr
    &    & &         { \bar{\tilde\eta}}^{-1ab ^+ } & \cr
  }
\nonu
\end{eqnarray}

\begin{eqnarray}
\matrix
{       &    &  &     B_{\mu \nu} &   &   & \cr
     &     &   \Psi^{(1)}_{\mu \nu}  &     &  &   \cr
     &    A^{(2)}_{\mu } &       &    & &   A^{(-2)}_{\mu } &  \cr
   R^{(3)} &          &  S^{(1)} &  & \bar \Psi^{(-1)}_{\mu}
           & &    \bar R^{(-3)}  \cr
      &   b_ {S^{(1)}} ^{(2)} &  & \Phi^{(0)} ,\bar\Phi ^{(0)}
              & & b_{\bar R^{(-3)}}^{(-2)}  \cr
       & & \eta^{(1)} &  &   &     &  &   \cr
  }
\nonu
\end{eqnarray}

\begin{eqnarray}
\matrix
{ \xi^ \mu   &     &    \bar \xi^ \mu \cr
     &    b^\mu }
\quad\quad\quad\quad\quad
\matrix
{ \O^ {ab  }  &   &        \bar \O^ {ab  } \cr
     &  b^{ab  }  & }
\label{spectrumfinal}
\end{eqnarray}
We will shortly see that the component $ \Psi^{( 1) }_{\mu\nu^+}$ 
of the topological ghost of the two-form $B_{\mu\nu}$  is gauge
fixed in an algebraic way, leaving  only $ \Psi^{( 1) }_{\mu\nu^-}$ as
a propagating field.  

Let us analyse the fields that remain 
after these eliminations. The fields
that will a play a r\^ole as the classical fields of supergravity are: 
\begin{eqnarray}
&&\mbox{bosons:}\ \  e^a_{\mu }, \sigma, B_{\mu\nu}, A^{(2)}_\mu,
A^{(-2)}_\mu \nn \\
&&\mbox{fermions:}\ \   
(\Psi^{(1)a}_\mu, \bar \Psi^{(-1)ab^-}_\mu, \bar
\Psi^{(-1)}_\mu), (\bar
\chi^{(-1)a}, \Psi^{( 1)ab^-},\chi^{( 1)})
\label{classic}
\end{eqnarray}
Indeed, these are nothing but   
the  fields of the $N=1,D=8$ supergravity multiplet, up to a twist.
The most striking point of our construction
is that the {\it topological} ghosts of ghosts of the two-form,
$A_\mu^{(2)}$ and
$A_\mu^{(-2)}$ can be interpreted as the propagating graviphotons
of $N=1,D=8$ supergravity.
The $Spin(7)$-scalar ghost of ghost
$\sigma$, which has ghost number zero, can be
interpreted as the dilaton.
The corresponding topological ghosts can be recognized as
the twisted version of the fermionic part of the spectrum.
Modulo some field redefinitions that we will discuss in detail in the
following section,
the twisted gravitino can be identified with the ghosts  
$(\Psi^{(1)a}_\mu, \bar \Psi^{(-1)ab^-}_\mu, \bar
\Psi^{(-1)}_\mu)$, and the twisted dilatino with 
$ (\bar \chi^{(-1)a },
 \Psi^{( 1)ab^-},\chi^{( 1)})$.
We remark that the BRST variation $sA_\mu^{(-2)}=\bar
\Psi _\mu^{(-1)}$ appears in the twisted gravitino. 
This is in agreement with the analysis of
\cite{BT1,BT2}, where we found that 
at least one  graviphoton with non zero ghost
number is needed both in 
four and eight dimensions in order to introduce $\bar
\Psi _\mu^{(-1)}$. 
As we will describe in the following section, the possibility 
of having $Spin(7)$ decompositions
allows one to do all relevant maps of tensors upon spinors. 

The fields of the topological multiplets not appearing in
eq.~(\ref{classic}) will be interpreted as the ordinary 
Faddeev--Popov ghosts and antighosts of the supergravity. 
The infinitesimal symmetry transformations of supergravity
will be deduced from the topological BRST equations.
After the eight-dimensional
untwisting, one gets from 
eq.~(\ref{spectrumfinal}) the propagating fields 
$(\Phi^{(2)a},  {\Phi}^{(0)ab -},\Phi^{(0)})$, 
$(\bar \Phi^{(-2)a},  {\t \Phi}^{(0)ab -},\bar \Phi^{(0)})$
and $(\bar \eta^{(-1)a},   \eta^{(1)ab ^-} , \eta^{(1)})$.
These can be interpreted respectively as 
the twisted version of the Faddeev--Popov spinorial 
ghosts and antighosts for local supersymmetry and the
corresponding Lagrange multipliers. 
The invariance of the untwisted theory therefore displays a
variation of the vector-spinor by a derivative of these ghosts:
this is presumably sufficient to ensure the full supersymmetry
of the untwisted action.
This is different from the case of topological Yang-Mills theory, where
the complete supersymmetric invariance of the untwisted action is not
automatic.

\section{The topological action and its correspondence with supergravity}

After having obtained the relevant field spectrum for the TQFT, the task
is of finding the topological gauge functions. 

A natural extension of the topological gauge function used in 
\cite{BT2} in the presence of the two-form field $B_{\mu\nu}$
is given by imposing the octonionic self-duality condition
\begin{equation}\label{octo}
\tilde\o^{ab}-\demi \O^{abcd} \tilde\o^{cd}=0,
\end{equation}
on the extended connection
\begin{equation}
\tilde\o^{ab}\equiv \o^{ab}- G^{ab}_c e^c  \ \ ,
\label{otilde}
\end{equation}
with torsion given by
\begin{equation}
\tilde T^a = \tilde D e^a = G^a_{bc}e^b e^c \ \ .
\label{torsion}
\end{equation}
The condition (\ref{octo}) has to be considered together with 
the gauge function
\begin{equation}
d\sigma + 2 *(\O\wedge G_3) =0
\label{g-sigma}
\end{equation}
involving the dilaton field.
We thus write the topological action
\begin{eqnarray}\label{top-act}
\L_{e,B}&=& s 
\Big[ 
\bar \Psi^{(-1)ac^-} (b ^{(0)cb^-} + \o^{cb^-}(e) - G_d^{cb^-} e^d)\V_{ab}  \cr
&&\ \ +\bar\chi^{(-1)a} \Big (b^{(0)}_a +\partial_a \sigma
+\Omega_{abcd}G_{bcd}  \Big  ) \Big],
\end{eqnarray}
where we introduced the compact notation for the volume forms 
${\cal V}_{a_1\ldots a_i}\equiv {1\over (8-i)!} 
\epsilon_{a_1\ldots a_8}e^{a_{(i+1)}}\ldots e^{a_8}$. 
This topological action, after integration on the Lagrange multipliers
$b^{(0)cb^-}$ and $b^{(0)}_a$, gives kinetic terms for the graviton, the
two-form and the dilaton of N=1 D=8 supergravity.

First of all, the Einstein Lagrangian, written with the curvature
$\tilde R^{ab}$ of the extended connection $\tilde\omega$, is equal to the
sum of the ordinary Einstein-Hilbert Lagrangian
plus the squared norm of the field-strength $G_3$
\begin{equation}
\tilde {\cal L} = \demi \tilde R^{ab}{\cal V}_{ab}
= \demi R^{ab}{\cal V}_{ab} - {1\over 2} G^{ab}_c G^{ab}_c 
{\cal V}.
\label{L-tilde}
\end{equation}
The Bianchi identity on
the field strength $G_3$ and (\ref{torsion}) imply that 
\begin{equation}
\tilde R^{ab}e_b e_a + \tilde T^a\tilde T^a = 0 \ \ .
\label{bi-t}
\end{equation}
By multiplying (\ref{bi-t}) by the self-dual four-form $\Omega$ we get 
\begin{equation}
\demi \O_{abcd}\tilde R^{cd} {\cal V}^{ab} =
\O_{mnpq}G^a_{mn}G^a_{pq} {\cal V}.
\label{ov}
\end{equation}
The relation (\ref{ov}), together with the $Spin(7)$
decomposition  
\begin{equation}\label{deco}
\tilde R^{ab}= \tilde R^{ab^+}+ \tilde R^{ab^-}, 
\quad \demi \O_{abcd} \tilde R^{ab}= \tilde R^{cd^+}-3 \tilde R^{cd^-} ,
\end{equation}
allows for the elimination of $\tilde R^{ab^+}$ in the Lagrangian
(\ref{L-tilde})
\begin{eqnarray}
\label{l-minus}
\tilde {\cal L} &=& \demi \tilde R^{ab}{\cal V}_{ab} = 
2\tilde R^{ab^-} {\cal V}_{ab} + {1\over 2}
\O_{abcd}G^{ab}_e G^{cd}_e {\cal V} \\
&=& -4 \tilde\o^{ac^-}\tilde\o^{cb^-}{\cal V}_{ab}
+ 2 \tilde\o^{ab^-}\tilde T^c \V_{abc}
+2d(\tilde\o^{ab^-}\V_{ab})
+{1\over 2}\O_{abcd}G^{ab}_e G^{cd}_e {\cal V}
\ \ ,
\nn
\end{eqnarray} 
where in the last identity we used that $\tilde R^{ab^-}=\tilde D\tilde\o^{ab^-}
-2\tilde\o^{ac^-}\tilde\o^{cb^-}$ and integrated by parts the term in
$\tD\ot^{ab^-}$. By comparing (\ref{l-minus}) with (\ref{L-tilde})
and using (\ref{torsion}) we can finally write the identity
\begin{equation}
 4 \tilde\o^{ac^-}\tilde\o^{cb^-}{\cal V}_{ab}=
{\cal L}_{EH} - {1\over 2}G^{ab}_c G^{ab}_c\V
+\O_{abcd}G^{ab}_f G^{cd}_f
+4 \o^{ab^-}G_{ab}^c\V_c
+2 d(\ot^{ab^-}\V_{ab}) \ \ ,
\label{omega^2}
\end{equation}
where we defined the Einstein-Hilbert action as
${\cal L}_{EH}\equiv \demi R^{ab}\V_{ba}$.
Let us consider now the square of the gauge function (\ref{g-sigma}).
By using the identity $\O_{abcd}\O^{afgh}=(6\delta_{bcd}^{[fgh]}
-9\O_{bc}^{[fg}\delta^{h]}_d )$ is easy to find that
\begin{equation}
\left(\pa_a\sigma + {1\over 3}\O_{abcd}G^{bcd}\right)^2 =
(\pa_a\sigma)^2 + {2\over 3}G^{ab}_c G^{ab}_c - 
\O_{abcd}G^{ab}_f G^{cd}_f + 
{2\over 3}\pa^a\left(\sigma\O_{abcd}G^{bcd}\right) \ \ .
\label{sigma^2}
\end{equation}
By summing (\ref{omega^2}) and (\ref{sigma^2})
(multiplied by the suitable volume form $\V$),
we finally get
\begin{eqnarray}
&& 4 \tilde\o^{ac^-}\tilde\o^{cb^-}{\cal V}_{ab} +
\left(\pa_a\sigma + {1\over 3}\O_{abcd}G^{bcd}\right)^2 \V 
\label{fund-id} \\
&&= {\cal L}_{EH} + \left[(\pa_a \sigma)^2 +  
{1\over 6}G^{ab}_c G^{ab}_c\right]\V
+ 4 \o^{ab^-}G_{ab}^c\V_c 
+ \mbox{boundary terms} \ \ .
\nn
\end{eqnarray}
This identity deserves some attention, since it shows 
that the sum of the 56 independent terms contained in $ |G _{abc}(B) |^2$,
plus those contained in the Einstein action can be obtained as a
topological gauge-fixing term, stemming from 64 Lagrange multipliers. 

We are now able to compare the topological terms (\ref{fund-id})
with the bosonic part of the action of N=1 D=8 supergravity.
In the spirit of topological field theory, we can restrict our attention
on the kinetic terms for the fields, which simplifies considerably
the comparison. In fact, the basic requirement 
on the gauge-fixing conditions is that they must give a good definition
for the propagators of the fields. Interaction terms can be then always
added as BRST exact terms in order to get agreement with the complete 
twisted supergravity action. 
The first three terms of the topological action (\ref{fund-id})
correctly reproduce the kinetic terms for the graviton, the two-form
and the dilaton of N=1 D=8 supergravity.
Concerning now the mixed term 
$ \o_c^{ab^-}(e)G_{abc}$ in (\ref{fund-id}),
it can be reduced in the quadratic approximation 
to 
\begin{equation}
\partial _\mu e^a_\mu \Omega_{abcd} G_{bcd} .
\label{gf}
\end{equation}
This can be done by imposing the $Spin(7)$ invariant Lorentz
gauge condition
$e^{[a}_\mu V^{b]^-\mu}=0$, where  $V^{b\mu}$ is an inverse 
8-bein chosen as a reference system.
\footnote{In the physicist language, this means that one uses 
the Lorentz invariance to impose that 7 of the components of 
$e^a_\mu$  vanish in a  $Spin(7)$ invariant way. Eventually, the rest
of the 21 Lorentz degrees of freedom can be used to enforce that
$e^a_\mu$ be a symmetrical
$8\times 8$ matrix, that is $e^a_\mu= V^{b\mu }  (\delta  ^{ab} +
h^{ab})$, where
$h^{ab}$ is symmetrical in $a$ and $b$. Then, there is a
one-to-one mapping between the metric and this matrix.}
The expression (\ref{gf}) can be absorbed  in the gauge-fixing
term for the reparametrization invariance
$s(\bar \xi^\mu \partial _\nu g_{\mu\nu})$.
We thus conclude that the topological action (\ref{fund-id})
coincides with the bosonic part of the supergravity action
in a given $Spin(7)$-invariant Lorentz gauge. This is not surprising
since the condition (\ref{octo}) leaves only a residual $Spin(7)$
symmetry group. 
It is remarkable that the kinetic energies of both the graviton and the
two-form stem from the gauge-fixing term (\ref{octo}) that only comes from the
topological freedom in the vielbein.

We now pass to the fermionic sector. The BRST variation
of the first line in (\ref{top-act})
gives part of the Rarita--Schwinger action, as in
\cite{BT2}. Notice that the condition $e^{[a}_\mu V^{b]^-\mu}=0$, which 
must be enforced by the BRST exact term $s  (\bar \O ^{[ab]^-}e^{[a}_\mu
V^{b]^-\mu})$,   yields the condition
$\O ^{[ab]^-}=-\Psi ^{[a}_\mu e ^{b]^-\mu}$, and ensures the $Spin(7)$
invariance of the fermionic part of the contribution of $\L_{e,B}$ to
the  Rarita--Schwinger action. There is actually a compensation between
the BRST variations of  $ \o _c^{ab^-}(e)$ and $G ^{ac^-}_c(B) $,
which is compulsory to enforce gauge invariance, since 
$\o   ^{ab^-}(e)$  transforms as a connection for Lorentz transformations 
with self-dual parameter $\O^{ab^-}$.

To determine the complete Rarita--Schwinger action, we must
add to $\L_{e,B}$ the following term, as in \cite{BT2}: 
\begin{eqnarray}
\L_{\bar  \Psi_\mu}=
s \Big[  \partial_{[\mu}A^{(-2)}_{\nu]^-} \Psi^{(1)a}_{[\mu}
e^a_{\nu]}\ \Big]
\end{eqnarray}
 Looking at the fermionic dependence, since $sA^{(-2)}_\mu=
\bar \Psi^{(-1)}_\mu$,  
one finds
that $ (\Psi^{(1)a}_\mu, \bar \Psi^{(-1) ab^-}_\mu, \bar
\Psi^{(-1)}_\mu)$,  can be identified as the twisted gravitino, with 8
chiral and 8 antichiral components, as in \cite{BT2}.

To determine  the propagation of $A^{(\pm 2)}_\mu$, we add:
\begin{eqnarray}
s\Big[\Psi^{(1)}_{\mu\nu} \pa_{[\mu} A^{(-2)}_{\nu]}  \Big]
=\pa_{[\mu} A^{(2)}_{\nu]}\pa_{[\mu} A^{(-2)}_{\nu]}
+\Psi^{(1)}_{\mu\nu}\pa_{[\mu}  \Psi^{(-1)}_{\nu]}
\end{eqnarray}
This identifies  $A^{(\pm 2)}_\mu$ as the two graviphotons of 
$N=1,D=8$ supergravity. Notice that among the two graviphotons
$A_\mu ^{(-2)}$ and 
$A_\mu ^{(2)}$, only the latter one has
a ``topological'' transformation, since
$sA_\mu^{(-2)}=\bar \Psi^{(-1)}_\mu$ while $sA_\mu^{(2)}=\partial _\mu
R^{(3)}$. This is in agreement with the twisted supergravity
transformations.

Part of the difficulty of this work was to 
understand the r\^ole of the ghost  field
$\Psi_{\mu\nu}^{(1)}$. On the one hand it is the field that generates by
its ghost of ghost symmetry the second graviphoton of the supergravity.
On the other hand,   in supergravity, the Kalb--Ramond two-form only
sees local supersymmetry through the gravitino and the dilatino, and it is
challenging to uncover this from  topological invariance. This leads to the
conclusion that not all the components of 
$\Psi_{\mu\nu}^{(1)}$ are independent propagating fields.  Only
$\Psi_{\mu\nu^-}^{(1)}$ survives as an independent field stemming
from the topological invariance of the two-form. 
To enforce the other components of $\Psi_{\mu\nu}^{(1)}$ in a BRST invariant 
way, we consider  the following
action, that exhausts the ghost of ghost symmetry in the Lorentz sector: 
\begin{eqnarray}\label{el}
\L_{\Psi^{(1)}_{\mu\nu},\tilde{\Phi}^{(2)ab}}&=&s\Big[ 
e^\mu_c e^\nu_d \  \bar{\tilde\Phi}^{(-2)cd^+}( {
\Psi}^{(1)}_{\mu\nu} -\Psi^{(1)a}_{[\mu}e^a_{\nu]})\Big]
\nn \\
&=&
e^\mu_c e^\nu_d \ 
\bar {\tilde \eta}^{(-1)cd^+}( { \Psi}_{\mu\nu}^{(1) }
-\Psi^{(1)a}_{[\mu} e^a_{\nu]})
+
\bar {\tilde \Phi}^{ab^+}( \partial _{[a}A^{(2)}_{ b]}
-\tilde \Phi  ^{( 2) }_{ab}+\ldots)\nonumber\\
\end{eqnarray}
This gauge-fixing allows one to identify $\Psi_{\mu\nu^+}^{(1)}=
\Psi^{(1)a}_{[\mu} e^a_{\nu]^+} $. If we define
\begin{equation}
\label{psib}
\Psi_{\mu\nu}^{(1)}=    \chi_{\mu\nu^-}^{(1)}
+\Psi^{(1)a}_{[\mu} e^a_{\nu]} 
\end{equation}
then $\chi_{\mu\nu^-}^{(1)}$ and $\chi^{(1)}=s\sigma$ are eight
fermionic variables that can be identified with a twisted
chiral component of the
dilatino. $\bar\chi^{(-1)\mu}$   determines by twist the other
chiral component. 
Moreover, Eq.(\ref{el}) 
accomplishes the  elimination of the ghost of ghost dependence in the Lorentz
sector, by yielding an algebraic equation of motion for 
$  {\tilde \Phi}^{( 2)ab^+}$ and $  \bar {\tilde \Phi}^{(-2)ab^+}$, in
a  $Spin(7)$ invariant way. 

Let us summarize the mapping between the fermionic degrees of freedom
of the topological action defined by (\ref{top-act})--(\ref{el})
and that of N=1 supergravity.
On a manifold with $Spin(7)$ holonomy there exist
a covariantly constant spinor (of norm one) $\varepsilon$~
\footnote{This spinor can be used to define the
self-dual four form as 
$ \O_{abcd}=\varepsilon^T \gamma_{abcd}\varepsilon$
\cite{bakasi}. }
, which can be used to redefine as in \cite{BT2}
the gravitino $(\lambda, \bar\lambda)$ and the
dilatino $(\chi, \bar\chi)$ of $N=1$, $D=8$ supergravity as
\footnote{Similar results concerning the twist of $N=1$, $D=8$
supergravity have been obtained by P. de Medeiros and B.Spence.}
\begin{eqnarray}
\lambda &=& \Psi^a \gamma_a \varepsilon \ \ , \nonumber\\ 
\bar\lambda &=& \bar\Psi \varepsilon + 
\bar\Psi^{ab^-}\gamma_{ab}\varepsilon\ ,
\label{twist-1} \\
\chi &=& \bar\chi^a \gamma^a \varepsilon \ \ , \nn\\
\bar\chi &=& \chi\varepsilon +
\chi^{ab^-}\gamma_{ab}\varepsilon \ \ .
\label{twist-2}
\end{eqnarray}
On the l.h.s. of (\ref{twist-1}), (\ref{twist-2}) $(\lambda, \bar\lambda)$
and $(\chi, \bar\chi)$ are spinors
of opposite chiralities. The eight-dimensional gamma matrices 
$\gamma_a$ acts on spinors of definite chirality.
Notice that the identification (\ref{psib}) implies the
appearance in the topological action (\ref{top-act}) of the mixed
kinetic terms $\bar\Psi^{ac^-}\partial_d\chi^{cb^-} e^d \V_{ab}$,
coming from the BRST-variation of the first line in (\ref{top-act}),
and $\bar\chi^a\O_{abcd}\partial_b\Psi_{cd}$, coming from the 
BRST-variation of the second line.
In order to
recover these terms from the twisted supergravity theory 
we have to impose the field redefinitions
$\Psi_b^a\rightarrow \Psi_b^a + \chi^{ab^-}$ and 
$\bar\Psi_c^{ab^-} \rightarrow \bar\Psi^{ab^-}_c + \delta^{[a}_c
\bar\chi^{b]^-}$. Moreover, since for the bosonic sector the
equivalence of the topological and supergravity actions
is valid only in a fixed $Spin(7)$-invariant Lorentz gauge,
we expect that also some gauge-fixing terms of the same kind
are involved in the comparison of the fermionic part.
From eq.~(\ref{twist-1}) and (\ref{twist-2})
we see that, modulo the above field redefinitions, the gravitino is mapped
to the fields $(\Psi^a, \bar\Psi, \bar\Psi^{ab^-})$ of the topological
model, while the dilatino is mapped to the fields
$(\bar\chi^a, \chi, \chi^{ab^-})$.

What we have found is interesting. The topological gauge functions are
such   that the BRST transformation in the effective topological action
for $e$ and $B$ is:
\begin{eqnarray}
s e^a_\mu &=& \Psi^{(1)a}_\mu +\Omega ^{ab} e^b_\mu 
+\ldots
\cr s B_{\mu\nu} &=& \Psi^{(1)a}_{[\mu} e^a_{\nu]}+
{\chi}^{(-1)}_{\mu\nu^-}+
\ldots  
\end{eqnarray}
Only the symmetrical part of $\Psi ^{(1)a}_\mu$ is involved in
$se^a_\mu$, while the Lorentz ghost $\Omega ^{ab}$ allows one to put to
zero the antisymmetrical part of $ e^a_\mu$ .
This explains  how the supersymmetric transformation law of the two-form  
in the supergravity framework can  be interpreted in a topological
way, using a suitable gauge function for the topological invariance.
What actually happens is that, when one twists the gravitino
$\Psi^\alpha_\mu$ into $\Psi^{(1)a}_\mu$, and   defines 
$\Psi^{(1)}_{ \mu\nu}=  e_{a\nu} \Psi^{(1)a}_\mu  $, 
then $\Psi^{(1)}_{\{\mu\nu\}}$ and $\Psi^{(1)}_{[\mu\nu]}$ are respectively the
topological ghosts of the two-form $B_{\mu\nu}$
and of the metric $g_{\mu\nu}$. Here, the mapping between the spinors
is not linear, since it involves various contractions by the vielbein.
This a important distinction with the case of the Yang--Mills
TQFT, where the mapping is a linear transformation. 

Some extra topological functions are needed in order to fix the 
symmetries of the gauge conditions used so far. These functions
exhaust the remaining fields of the BRST
topological multiplets. Let us briefly sketch them.
To take care of the gauge invariance of $\bar \Psi_\mu ^{(-1)ab^-}$,
which follows from the gauge functions, we redefine $b^{(0)ab^-}_\mu \to
b^{(0)ab^-}_\mu +\partial _\mu \t \Phi ^{(0)ab^-}$.
Then, to  gauge fix the local supersymmetry, which
pops up as the gauge invariance of the topological ghosts,  we add, as in
\cite{BT1,BT2}
\begin{eqnarray}
\label{gfghost}
\L_{ghosts}=
s\Big[\sqrt{g}(\bar \Phi ^{(-2)a}   D_\mu   \Psi ^{{(1)a}  }_\mu+
{\Phi}^{(0)ab^-}  D_\mu   \bar \Psi ^{(-1)ab^-}_\mu
+\bar \Phi ^{(0) } \pa_\mu   \bar \Psi^{(-1) }_\mu)\Big]
\end{eqnarray}
The r\^ole of this redefinition of $b^{ab^-}$ has been
analysed in \cite{BT1, BT2}. It ensures the propagation of
the field $\t \Phi ^{(0)ab^-}$.
The expression of  this  action identifies    $(\Phi^{(2)a}, 
{\Phi}^{(0)ab -},\Phi^{(0)})$,
$(\bar \Phi^{(-2)a},  {\t \Phi}^{(0)ab -},\bar \Phi^{(0)})$ 
and $(\bar \eta^{(-1)a},   \eta^{(1)ab ^-} , \eta^{(1)})$ 
as the twisted version of the  Faddeev--Popov spinorial 
ghost and antighosts for local supersymmetry, and their fermionic   
Lagrange multipliers, respectively.

We also  use the topological gauge freedom 
of the spin connection
$\o^{ab }$ to eliminate  this field  
in terms of $e$, by mean of the term 
\begin{eqnarray}
s\Big [ \bar {\tilde \Psi}^{(-1)ab} e_b \wedge * T^a \Big]
\end{eqnarray}  
where $*T$ is the Hodge dual of the torsion $T=de+\o\w e $.
This gauge fixing, which
can be improved by changing $ T^a\to T^a +G^a_{bc} e^be^c$ , also
trivially  eliminates  the dependence of the action upon the topological
Lorentz ghosts
 $ {\tilde \Psi}^{(1) ab}_\mu$ and  $\bar {\tilde \Psi}^{(-1)ab}_\mu$,
which disappear by their algebraic equations of motion.
One must recognize that introducing the Lorentz symmetry is extremely
useful, although most of its ingredients are eventually
eliminated.

As for the fields $   S^{(1)}$ and $R^{(-3)}$, they are used to
fix the ordinary gauge symmetry of 
$A^{(-2)}$ and $A^{(2)}$ by the action
\begin{eqnarray}
s\Big [   S^{ (1)}  \partial_\mu A^{(-2)}_\mu  +\bar R^{(-3)}\partial_\mu
A^{(2)}_\mu  \Big ]
\end{eqnarray}
To impose that the vielbein is a symmetrical matrix, and eliminate the 
$\Omega$ and $\bar \O$ dependence, we just add :
\begin{eqnarray}
s \Big[   \bar \Omega  ^{ab^+}   e^{b}_\mu V ^{\mu a} \Big]
=\bar \Omega  ^{ab^+}   (\Omega  ^{ab}  +\ldots)
+b  ^{ab^+}  
e^{b}_\mu V ^{\mu a},
\end{eqnarray}
keeping in mind that we had already used a term
$s \Big[   \bar \Omega  ^{ab^-}   e^{b}_\mu V ^{\mu a} \Big]$.  After
expansion, and a few field redefinitions, one gets that
$ \Omega ^{ab}$ and $ \bar \Omega ^{ab}$  are  eliminated by Gaussian
integration.  As said above, this necessitates the introduction of an 
inverse vielbein $V^{a\mu}$ as background.

Last of the last, we must add
\begin{eqnarray}
s\Big [\bar\xi ^{-1\mu }\partial_\nu g_{\mu\nu}  \Big ]
\end{eqnarray}
to fix the reparametrization invariance.

The gauge-fixing of the  gauge symmetries of  
topological gauge functions could have been done in a much more refined
way, using the technology of equivariant cohomology\footnote {As
remarked in the previous section, the
equivariance is with respect to reparametrizations, $Spin(7)\subset
SO(8)$ Lorentz invariance and two-form gauge symmetry.}.
This is a technicality that we have chosen not
to present here. It would     distract us from our main result,  
that is, we have finally end up our task of building a TQFT for the
Kalb--Ramond field
$B_{\mu\nu}$  within the context of
topological gravity. The result is that the standard TQFT
procedure has lead us to a twisted version of $N=1, D=
8$ supergravity.

It is worth noticing that the topological gauge functions
on the extended connection (\ref{octo}) and 
on the dilaton (\ref{g-sigma}) are the same appearing in the
octonionic superstring equations \cite{stro}.
By coupling our model with a non abelian
topological Yang-Mills theory, as it is defined 
in~\cite{bakasi,Acharya:1997gp},
one could thus obtain a topological theory which effectively
describes the transverse properties of the octonionic
superstring.

We believe that it will be interesting to study the possible 
dimensional reductions of the 8-dimensional topological gravity. 
This idea  has already  proven to be quite useful in the simpler case of 
the topological Yang--Mills theory.
The reduction to seven dimensions of the octonionic
self-duality conditions on the spin connection
is of relevance for
manifolds with (weak) $G_2$ holonomy     
\cite{Bilal}. The study of the dimensional reduction
of the generalized self-duality conditions (\ref{octo})
and (\ref{g-sigma}) could give a more general description
of manifolds with $G_2$-structure.

Moreover, one may investigate the possibility
of getting a generalization of the Seiberg--Witten   
theory involving gravity.  Generalizing the flat space analysis 
of \cite{bakasi}, this theory could be derived as an adequate dimensional
reduction in four dimensions of the 8-dimensional 
topological gravity. Further dimensional reductions can give 
interesting models  in 2 and 0 dimensions.

\vspace{1.cm}

{\bf Acknowledgments}: 

\vspace{.5cm}
\noindent 
We thank P. de Medeiros and B. Spence for useful discussions.
A.T.~is supported by a Marie Curie fellowship under contract
N$^o$ HPMF-CT-2001-01504.

\end{document}